\documentclass[showpacs,preprintnumbers,single,amsmath,amssymb,floatfix,elsart,prd]{revtex4}

\usepackage{graphicx}
\usepackage{dcolumn}
\usepackage{bm}
\usepackage{hyperref}

\begin{document}

\title{Relativistic compact stars in the Kuchowicz spacetime}

\author{\bf K N Singh$^{1,2}$\footnote{Corresponding author: ntnphy@gmail.com} , F Rahaman$^2$, N Pradhan$^1$ and N Pant$^3$}
\affiliation{$^1$Department  of  Physics,  National  Defence  Academy,Khadakwasla, Pune-411023,  India\\
$^2$Faculty Council of Science, Jadavpur  University,  Kolkata-700032,  India\\
$^3$Department of Mathematics, National Defence Academy, Khadakwasla, Pune-411023, India}

\date{\today}

\begin{abstract}
{\bf Abstract: }We present an anisotropic charged analogue of Kuchowicz (1971) solution  of the general relativistic field equations in curvature coordinates by using simple form of electric intensity $E$ and pressure anisotropy factor $\Delta$ that involve charge parameter $K$ and anisotropy parameter $\alpha$ respectively. Our solution is well behaved in all respects for all values of $X$ ( $X$ is related to the radius of the star )   lying in the range $0< X \le 0.6$, $\alpha$ lying in the range $0 \le \alpha \le 1.3$, $K$ lying in the range $0< K \le 1.75$ and Schwarzschild compactness parameter ``$u$" lying in the range $0< u \le 0.338$. Since our solution is well behaved for a wide range of the parameters, we can model many different types of ultra-cold compact stars like quark stars and neutron stars. We present some models of super dense quark stars and neutron stars corresponding to $X=0.2,~\alpha=0.2$ and $K=0.5$ for which  $u_{max}=0.15$. By assuming surface density $\rho_b=4.6888\times 10^{14}~ g/cc$  the mass and radius are  $0.955 M_\odot$ and $9.439 km$ respectively.   For $\rho_b=2.7\times 10^{14}~ g/cc$  the mass and radius are  $1.259 M_\odot$ and $12.439 km$  respectively and for $\rho_b=2\times 10^{14}~ g/cc$ the mass and radius are  $1.463 M_\odot$ and $14.453 km$  respectively. It is also shown that inclusion of more electric charge and anisotropy enhances the static stable configuration under radial perturbations. The $M-R$ graph suggests that the maximum mass of the configuration depends on the surface density {\bf i.e. with the increase of surface density} the maximum mass and corresponding radius decrease. This may be because of existence of exotic matters at higher densities that soften the EoSs.\\


\end{abstract}

\maketitle

\section*{ 1. Introduction}\label{Sec1}

Ever since Oppenheimer and Volkoff \cite{opp39} analyzed and determined the maximum mass of compact astrophysical objects solving Einstein's field equations, many researchers are inspired to discover more solutions by introducing anisotropy, charge, rotation, magnetic and electric field, equation of state (EoS) etc. However, due to the non-linearity and highly coupled nature of the field equations it is difficult to find physically possible solutions. Even though, many astrophysical objects such as neutron star (bound by gravity) or self-bound strange quark star (bound by the strong interaction) where one needs to include relativistic gravity to reconsider the EoS.

In the recent past decades the observational data put very strong constraints in choosing EoS to represent compact stars. However, due to its complex composition and lack of knowledge, it is always a difficult task to determine the exact EoS. Furthermore, looking into the internal structure of the space-time we can always construct the analytical EoS using field equations. This method can be proceeded by several ways assuming particular form of metric potentials, EoS, anisotropy, charge density, mass function, density profile, killing symmetry etc.

Compact stars are formed at the end of main sequence stars whose masses are sufficient to form white dwarfs or neutron stars or black holes. In the entire evolution, the magnetic flux is always conserved and termed as ``flux freezing". Due to flux freezing, the surface magnetic field of compact stars are very high $\sim 10^{12}-10^{14}$ G that can lead to anisotropic in the matter distribution \cite{we}. Also, during the stages of evolution the angular momentum remains conserved leading to very high rotational kinetic energy of compact stars. Due to the rotational motion of the magnetic moment, it leads to emission of radio waves and observed as pulsars \cite{pac68,gol68}.

Various theories have also been suggested that the matter distribution inside the compact stars need not be always isotropic. In better realistic picture, anisotropic distribution is more likely and it can be due to presence of solid core or type-III A super-fluid \cite{kip90}, electric charge \cite{uso04}, phase transitions \cite{sok80}, meson condensation \cite{saw72}, slow rotation \cite{her97}, a mixture of two gases \cite{let80}. The stress tensor of anisotropic matter may also be expressed as two perfect fluids, or a perfect fluid and a null fluid, or two null fluids \cite{let80,let81,let82,bay82}.

Ivanov \cite{iva02} has shown for the first time that inclusion of charge in perfect fluid inhibits the growth of space time curvature to avoid singularities. Bonnor \cite{bon65} pointed out that a dust distribution of arbitrarily large mass and small radius can remain in equilibrium against the pull of gravity by a repulsive force produced by a small amount of charge. Thus it is desirable to study the implications of Einstein-Maxwell field equations with reference to the general relativistic prediction of gravitational collapse. For these purposes anisotropic and charged fluid ball models are required. The external field of such ball has {\bf to be} matched with Reissner-N$\ddot{o}$rdstrom solution.

Dev and Gleiser \cite{dev03} demonstrated that pressure anisotropy affects the physical properties, stability and structure of stellar matter. The stability of stellar bodies is improved for positive measure of anisotropy when compared to configurations of isotropic stellar objects. Dev and Gleiser \cite{dev03}, Gleiser and Dev \cite{gle04} showed that the presence of anisotropic pressures in charged matter enhances the stability of the configuration under radial adiabatic perturbations as compared to isotropic matter. Malaver has proposed some strange quark star models with anisotropy in the framework general relativity theory \cite{mal1}-\cite{mal4}. Pant et al. \cite{mal5} have found new exact solutions of the field equations for anisotropic neutral fluid in isotropic coordinates. Many authors have also presented many solutions on charge and /or anisotropic solutions \cite{sin15}-\cite{pan12}.

In our solution, we choose seed solution of Kuchowicz \cite{kuc68} and {\bf find} a solution by assuming appropriate functional form of charge parameter \cite{pan12} as well as anisotropic parameter in such a way that the obtained solution is well behaved in all respects.

\begin{figure}[t]
\centering
\resizebox{0.5\hsize}{!}{\includegraphics[width=4cm,height=3cm]{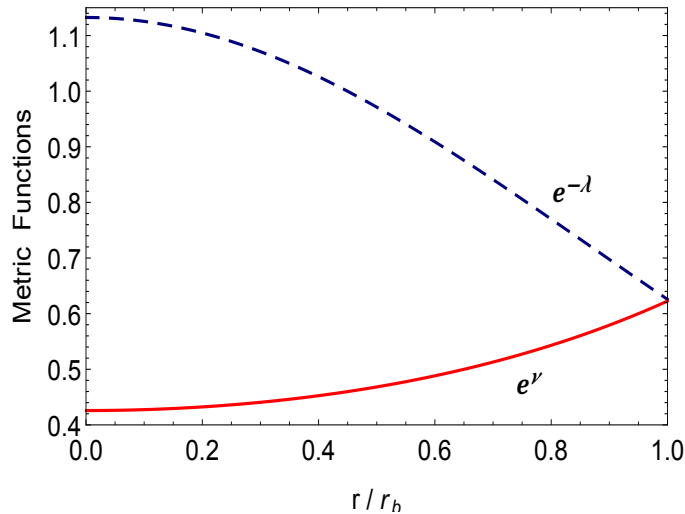}}
\caption{Metric potnetials are plotted against $r$ inside the stellar interior by taking $X=0.32,~K=0.99$ and $\alpha=0.2$.}\label{fig1}
\end{figure}

\begin{figure}[t]
\centering
\resizebox{0.6\hsize}{!}{\includegraphics*{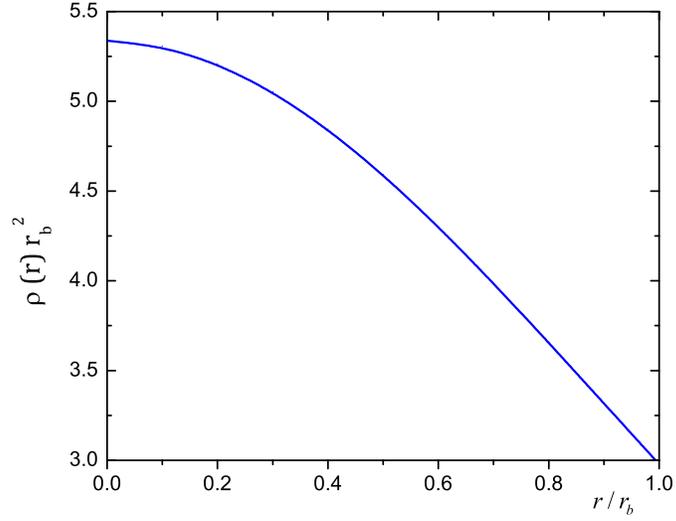}}
\caption{Matter density is plotted against $r$.}\label{fig2}
\end{figure}

\begin{figure}[t]
\centering
\resizebox{0.6\hsize}{!}{\includegraphics*{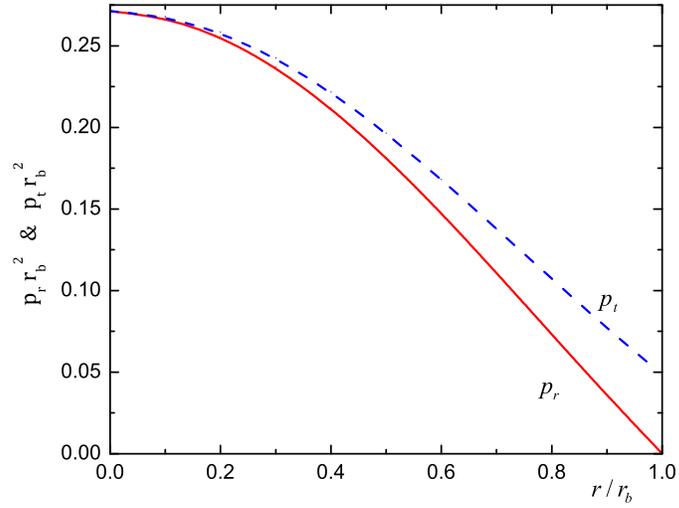}}
\caption{Pressures are plotted against $r$.}\label{fig3}
\end{figure}

\begin{figure}[t]
\centering
\resizebox{0.6\hsize}{!}{\includegraphics*{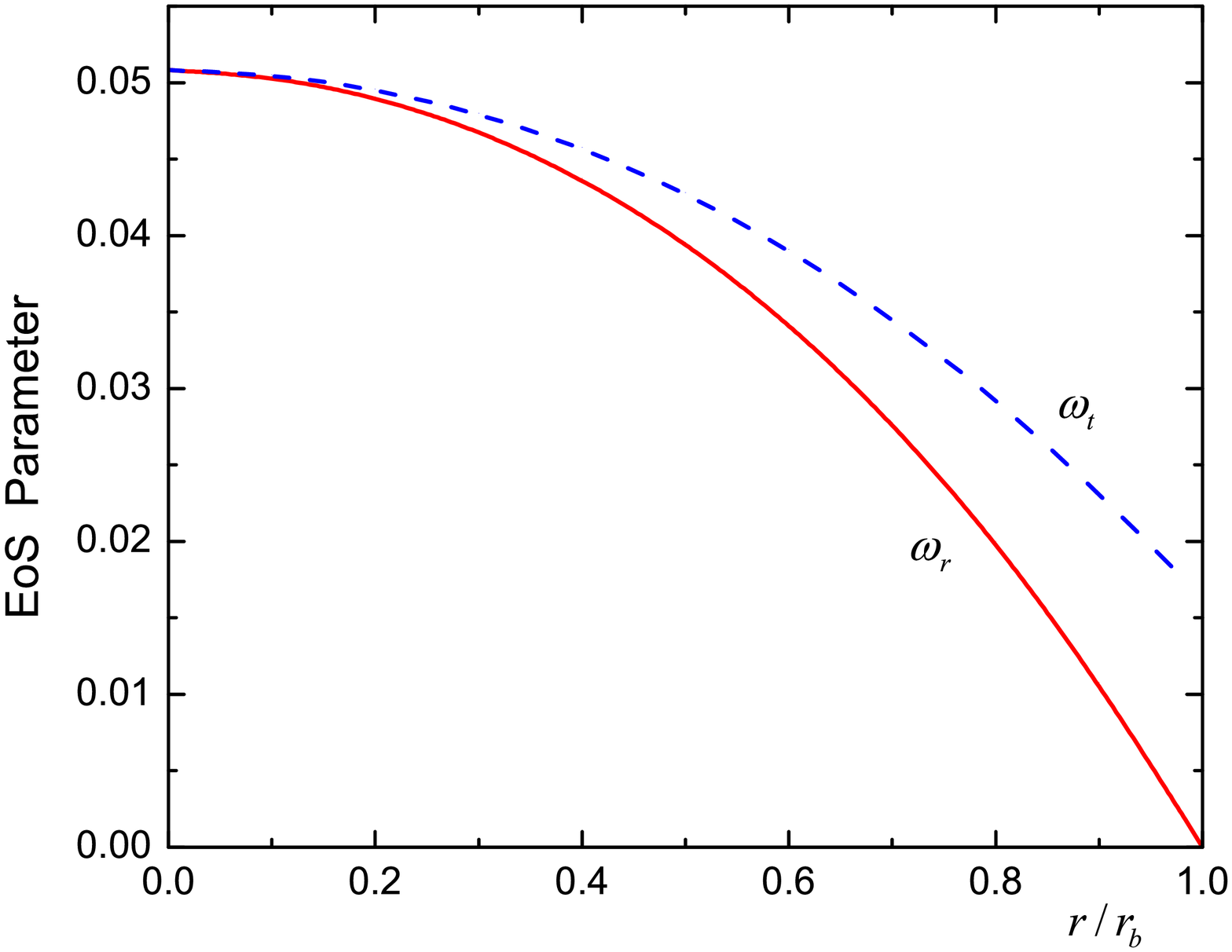}}
\caption{Equation of state parameters are plotted against $r$.}\label{fig4}
\end{figure}

\section*{2. Basic field equations}
To describe the interior of a static and spherically symmetric object the line element can be taken in canonical co-ordinate as,
\begin{equation}
ds^{2}=-e^{\nu(r)}dt^{2}+e^{\lambda(r)}dr^{2}+r^{2}\left(d\theta^{2}+\sin^{2}\theta d\phi^{2} \right), \label{met}
\end{equation}
where $\nu$ and $\lambda$ are functions of the radial coordinate `$r$' only.\\

The field equations for charged anisotropic fluid distribution can be written as
\begin{eqnarray}
R^\mu_\nu-{1\over 2}g^\mu_\nu R &=& -{8\pi} \bigg[(p_t +\rho c^2)v^\mu v_\nu-p_t g^\mu_\nu+(p_r-p_t) \chi_\nu \chi^\mu+{1 \over 4\pi} \bigg(-F^{\mu \lambda} F_{\nu \lambda}+{1 \over 4}~g^\mu_\nu F^{\alpha \beta} F_{\alpha \beta}\bigg) \bigg] \label{fil}\\
-4\pi J^\mu &=& {1 \over \sqrt{-g}} ~\partial_\beta \left( \sqrt{-g} ~F^{\mu \beta}\right)\\
0 &=& \partial_\beta F_{\mu \nu}+\partial_\mu F_{\nu \beta}+\partial_\nu F_{\beta \mu}.
\end{eqnarray}
where the symbols have their usual meanings.

\begin{figure}[t]
\centering
\resizebox{0.5\hsize}{!}{\includegraphics[width=4cm,height=3cm]{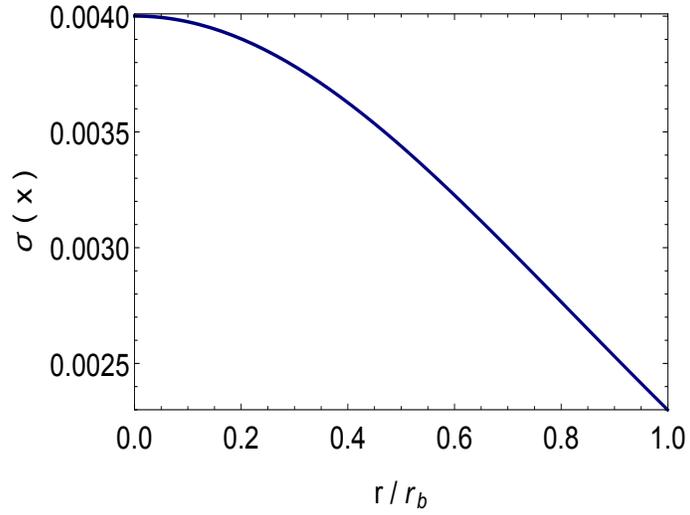}}
\caption{Invariant charge density is plotted against $r$.}\label{fig5}
\end{figure}

\begin{figure}[t]
\centering
\resizebox{0.6\hsize}{!}{\includegraphics*{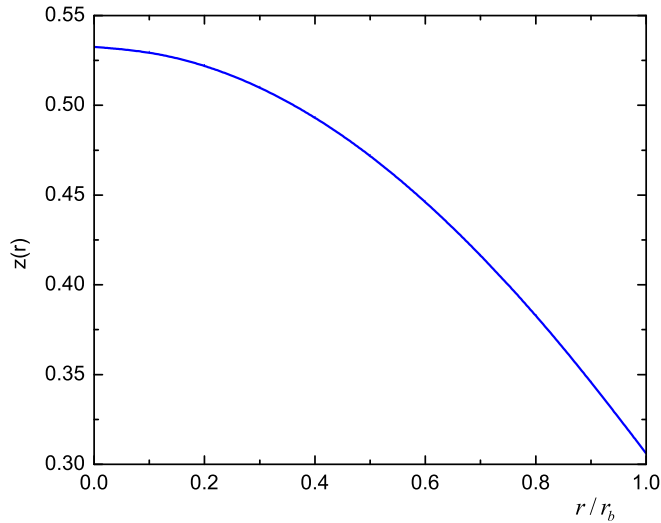}}
\caption{Red-shift is plotted against $r$.}\label{fig7}
\end{figure}

The Einstein-Maxwell's field equations (\ref{fil}) for the chosen spacetime (\ref{met}) reduce to
\begin{eqnarray}
\frac{1-e^{-\lambda}}{r^{2}}+\frac{e^{-\lambda}\lambda'}{r} &=& 8\pi\rho+E^{2} \label{dens}\\
\frac{e^{-\lambda}-1}{r^{2}}+\frac{e^{-\lambda}\nu'}{r} &=& 8\pi p_{r}-E^{2} \label{prs}\\
e^{-\lambda}\left(\frac{\nu''}{2}+\frac{\nu'^{2}}{4}-\frac{\nu'\lambda'}{4}+\frac{\nu'-\lambda'}{2r} \right) &=& 8\pi p_t+E^{2} \label{prt}\\
{e^{-\lambda/2} \over 4\pi r^2} \Big(r^2E\Big)' &=& \sigma(r)	\label{char}
\end{eqnarray}
where $\sigma(r)$ is the charge density and $E=q(r)/r^2$  the electric field intensity at the interior.\\

Using (\ref{prs}) and (\ref{prt}) we get
\begin{eqnarray}
&& e^{-\lambda} \left( {\nu'' \over 2}+{\nu'^2 \over 4}-{\nu' \over 2r}-{1 \over r^2}\right)-e^{-\lambda} \lambda' \left({\nu' \over 4}+{1 \over 2r} \right) +{1 \over r^2}-\Delta -2E^2 = 0 \label{eqn}
\end{eqnarray}

\begin{figure}[t]
\centering
\resizebox{0.5 \hsize}{!}{\includegraphics[width=4cm,height=3cm]{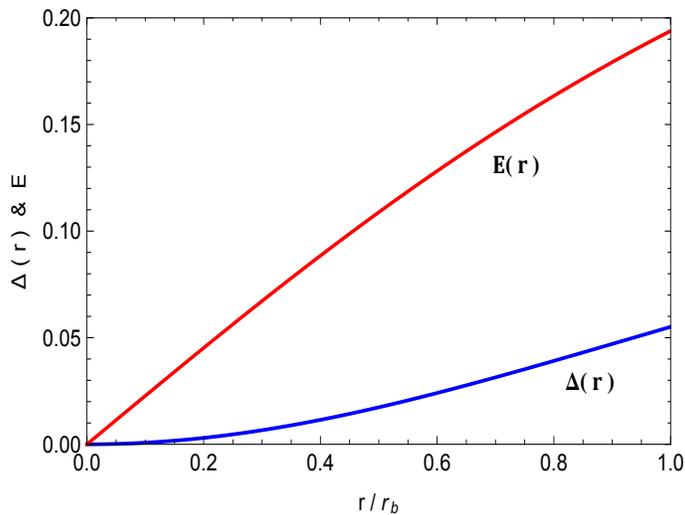}}
\caption{Electric field and anisotropy are plotted against $r$.}\label{fig6}
\end{figure}

\begin{figure}[t]
\centering
\resizebox{0.5\hsize}{!}{\includegraphics[width=4cm,height=3cm]{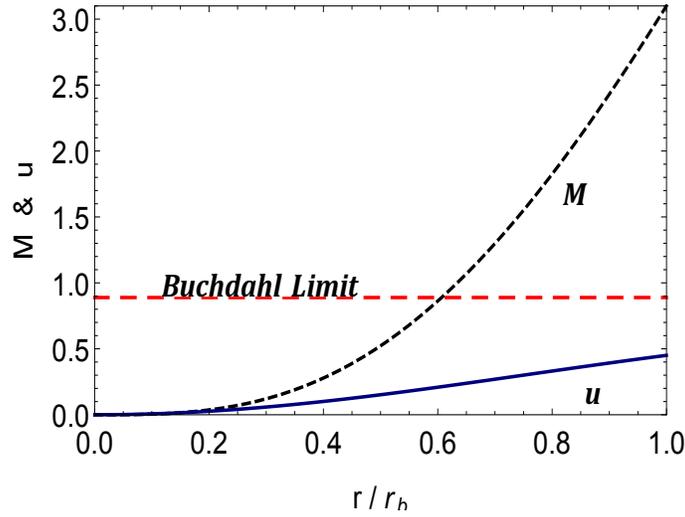}}
\caption{Variations of mass and compactness factor with radial coordinate.}\label{fig12}
\end{figure}

\begin{figure}[t]
\centering
\resizebox{0.5\hsize}{!}{\includegraphics[width=4cm,height=3cm]{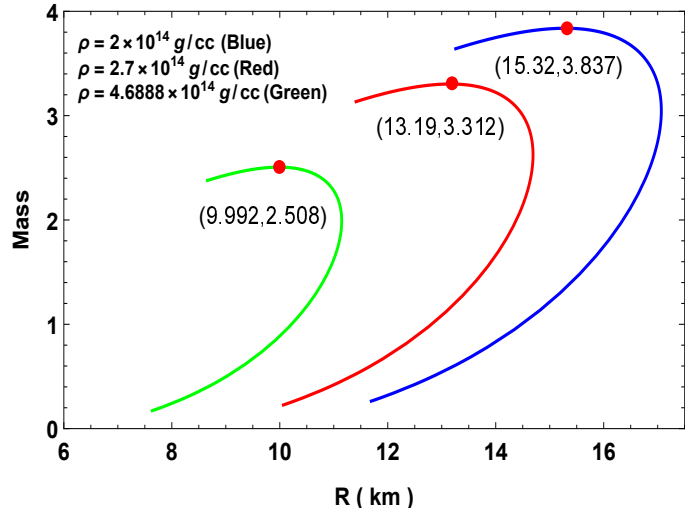}}
\caption{M-R graph for various surface densities.}\label{mrn}
\end{figure}

with $\Delta = 8\pi (p_t - p_r)$ is the measure of anisotropy.\\

Assuming $x=Cr^2$ and $y=e^{-\lambda}$, (\ref{eqn}) transforms into
\begin{eqnarray}
(1+x\dot{\nu}) {dy \over dx}+\left(2x\ddot{\nu}+x\dot{\nu}^2-{1 \over x} \right)y+{1 \over x}-{\Delta \over C}-{2E^2 \over C}=0. \nonumber \\ \label{eqn1}
\end{eqnarray}
Here  $\dot{\nu}=d\nu/dx$. The above equation includes four unknown variables, hence three of them are needed to ansatz.

\section*{3. Anisotropic charged Kuchowicz spacetime}

Assuming the $g_{tt}$ metric potential of the form of Kuchowicz \cite{kuc68}, particular form of $\Delta$ and $E$ as
\begin{eqnarray}
e^\nu= B e^x~,~~\Delta = {C\alpha x \over 1+x}~,~~{2E^2 \over C} = {2Cq^2 \over x^2} = {K x \over 1+x}. \label{assum}
\end{eqnarray}
where $K,~\alpha , ~C,~B$ are non-zero positive constants. Here we have chosen the form of $\Delta$ and $E$ given in (\ref{assum}) so that they vanish at the center, monotonically increasing function of $r$ and most importantly, Eq. (\ref{eqn1}) must be integrable. Since the center of the core is highly dense, the mean free path is expected to be very small and the chance of happening electron capture i.e. $p+e^- \rightarrow n+\nu_e$ is very high leaving the center electrically neutral. However, as we move towards the surface the density decreases and hence the mean free path increases. Therefore, free charged particles can exist with less probability of recombination leading to higher electric field intensity. In the similar way, the highly compact core makes the pressure isotropy at the very center and as moving outward the increase is electric field along radial direction makes the pressure more anisotropic on the surface.\\

On using (\ref{assum}) in (\ref{eqn1}) we get
\begin{eqnarray}
{dy \over dx} + {x-1 \over x}~y+{1+x-x^2(\alpha+K) \over x(1+x)^2}=0
\end{eqnarray}
whose solution can be written as
\begin{eqnarray}
y = e^{-\lambda} &=& 1+Ax~e^{-x}-{(\alpha+K)x \over 1+x}+{\alpha+K-1 \over e^{1+x}} \text{Ei}(1+x)
\end{eqnarray}
where $A$ is the constant of integration. Here $\text{Ei}(\kappa)$, the exponential integral for real non-zero $\kappa$ defined as
\begin{equation}
\text{Ei}(\kappa) = - \int_{-\kappa}^\infty {e^t \over t}~dt.
\end{equation}
The two metric functions are plotted in Fig. \ref{fig1}.

The physical parameters can be written as
\begin{eqnarray}
{{8\pi} \rho \over C}  &=& {4x+4x^2+2\alpha(3-x-2x^2)+K(6-3x-5x^2) \over 2(1+x)^2} + (2x-3)Ae^{-x} + {(2x-3)(K+\alpha-1) \over e^{1+x}}~\text{Ei}(1+x)\label{r1}\\
{8\pi p_r  \over C} &=& {4+4x-\alpha(2+4x)-K(2+3x) \over 2(1+x)} + (1+2x) Ae^{-x} + {(1+2x)(K+\alpha-1) \over e^{1+x}}~\text{Ei}(1+x) \\
{{8\pi } p_t \over C}  &=& {4+4x-\alpha(2+2x)-K(2+3x) \over 2(1+x)} + (1+2x) Ae^{-x} + {(1+2x)(K+\alpha-1) \over e^{1+x}}~\text{Ei}(1+x)  \label{r2}\\
\sigma(x) &=& \frac{\sqrt{C} (2 x+3) }{ 8 \sqrt{2} \pi  K x^{3/2}} \left(\frac{C K x}{x+1}\right)^{3/2} \Big[A e^{-x} x+e^{-x-1} \text{Ei}(x+1) (\alpha +K-1)-\frac{x (\alpha +K)}{x+1}+1 \Big]^{1/2}.\label{r3}
\end{eqnarray}

All this physical quantities are plotted in Figs. \ref{fig2}-\ref{fig5}.

The corresponding gradients can be written as
\begin{eqnarray}
{{8\pi} \dot{\rho} \over C}  &=& {10+2x-4x^2-11K-10\alpha +4Kx+4\alpha x \over 2(1+x)^2} +{4x(K+\alpha) \over (1+x)^2}+(5-2x) \bigg[ Ae^{-x}-{K+\alpha \over 1+x} \nonumber \\
&& +{K+\alpha-1 \over e^{1+x}}~\text{Ei}(1+x) \bigg] \label{dro}\\
{{8\pi} \dot{p}_r \over C}  &=& {4x^2-2-6x+3K+2\alpha +4Kx+4\alpha x \over 2(1+x)^2} +(1-2x) \bigg[Ae^{-x}-{K+\alpha \over 1+x} +{K+\alpha-1 \over e^{1+x}} \text{Ei}(1+x)\bigg] \label{dpre}\\
{{8\pi \dot{p}_t} \over C}   &=& {4x^2-2-6x+3K+4\alpha +4Kx+4\alpha x \over 2(1+x)^2} +(1-2x) \bigg[Ae^{-x}-{K+\alpha \over 1+x} +{K+\alpha-1 \over e^{1+x}} \text{Ei}(1+x)\bigg]. \label{dprt}
\end{eqnarray}

Here $x = X (r^2/r_b^2)$ and $X=Cr_b^2$ with $r_b$ as the radius of the configuration.

\section*{4. Non-singular nature of the solution}

The physical validity of the solution is necessary to {\bf  be checked} via central values of the physical quantities. The central values of pressure and density can be written as
\begin{eqnarray}
{{8\pi} p_{r0} \over C}  &=& {{8\pi} p_{t 0} \over C}  = 1+(K+\alpha -1) \left[{\text{Ei}(1) \over e}-1 \right]+A  >0, \label{rhc}\\
{{8\pi} \rho_0 \over C}  &=& 3(K+\alpha-A)-\left[{3(K+\alpha-1)\text{Ei}(1) \over e} \right]  >0. \nonumber \\ \label{pc} \end{eqnarray}

For any physical fluid distribution the Zeldovich's criterion is necessary to {\bf satisfied} i.e. $p_{r0}/ \rho_0 \le 1$ which implies
\begin{eqnarray}
A \le {e(2K+2\alpha-1)-2(K+\alpha-1) \text{Ei}(1) \over 2e}. \label{zel}
\end{eqnarray}

The required range of $A$ can be found from (\ref{rhc}) and (\ref{zel}) as
\begin{eqnarray}
K+\alpha-2{(K+\alpha-1)\text{Ei}(1) \over e} < A \le {e(2K+2\alpha-1)-2(K+\alpha-1) \text{Ei}(1) \over 2e}.
\end{eqnarray}

Now the expression for gravitational red-shift is given as
\begin{equation}
z(x) = e^{-\nu(x)/2}-1={1 \over \sqrt{B}}~e^{-x/2}-1. \label{reds}
\end{equation}
Variation of red-shift is plotted against radial coordinate in Fig. \ref{fig7}. Since the central value of gravitational red-shift has to be non-zero, positive finite, we have  $0<\sqrt{B}<1$. Differentiating (\ref{reds}) w.r.t. $x$ we get
\begin{equation}
\left[{dz \over dx}\right]_{x=0} = -{1 \over 2\sqrt{B}} <0
\end{equation}
which implies that the gravitation red-shift is maximum at the center and decreases outward.

Similarly the derivatives of electric field and anisotropy at the center are given as

\begin{equation}
\left[{d \over dx} {E^2 \over C}\right]_{x=0} = {K \over 2} >0 ~~~\mbox{and}~~~\left[{d \Delta \over dx}\right]_{x=0} = C \alpha >0
\end{equation}
signify that electric field and anisotropy are minimum (i.e. zero) at the center and monotonically increasing outward (Fig. \ref{fig6}).

\section*{5. Boundary Conditions and determination constants}

We match our interior space-time to the exterior $Reissner- N\ddot{o}rdstrom$ line element given by
\begin{eqnarray}
ds^{2} &=& -\left(1-\frac{2m}{r}+\frac{q^2}{r^{2}}\right)dt^{2}+\left(1-\frac{2m}{r}+\frac{q^2}{r^{2}}\right)^{-1}dr^{2} +r^{2}(d\theta^{2}+\sin^{2}\theta d\phi^{2})
\end{eqnarray}
with the radial coordinate $r$ must be greater than $m+\sqrt{m^{2}-q^2}$ so that it doesn't form a black hole.

Using the continuity of the metric coefficient $e^{\nu}$ and $e^{\lambda}$ across the boundary ($r=r_b$)  we get the following equations
\begin{eqnarray}
1-\frac{2M}{r_b}+\frac{q^2(r_b)}{r_b^{2}} &=& e^{\nu_b} \label{b1}\\
\left(1-\frac{2M}{r_b}+\frac{q^2(r_b)}{r_b^{2}}\right)^{-1} &=& e^{\lambda_b}   \label{b2}\\
p_r(r=r_b) &=& 0. \label{b3}
\end{eqnarray}

\begin{figure}[t]
\centering
\resizebox{0.7\hsize}{!}{\includegraphics*{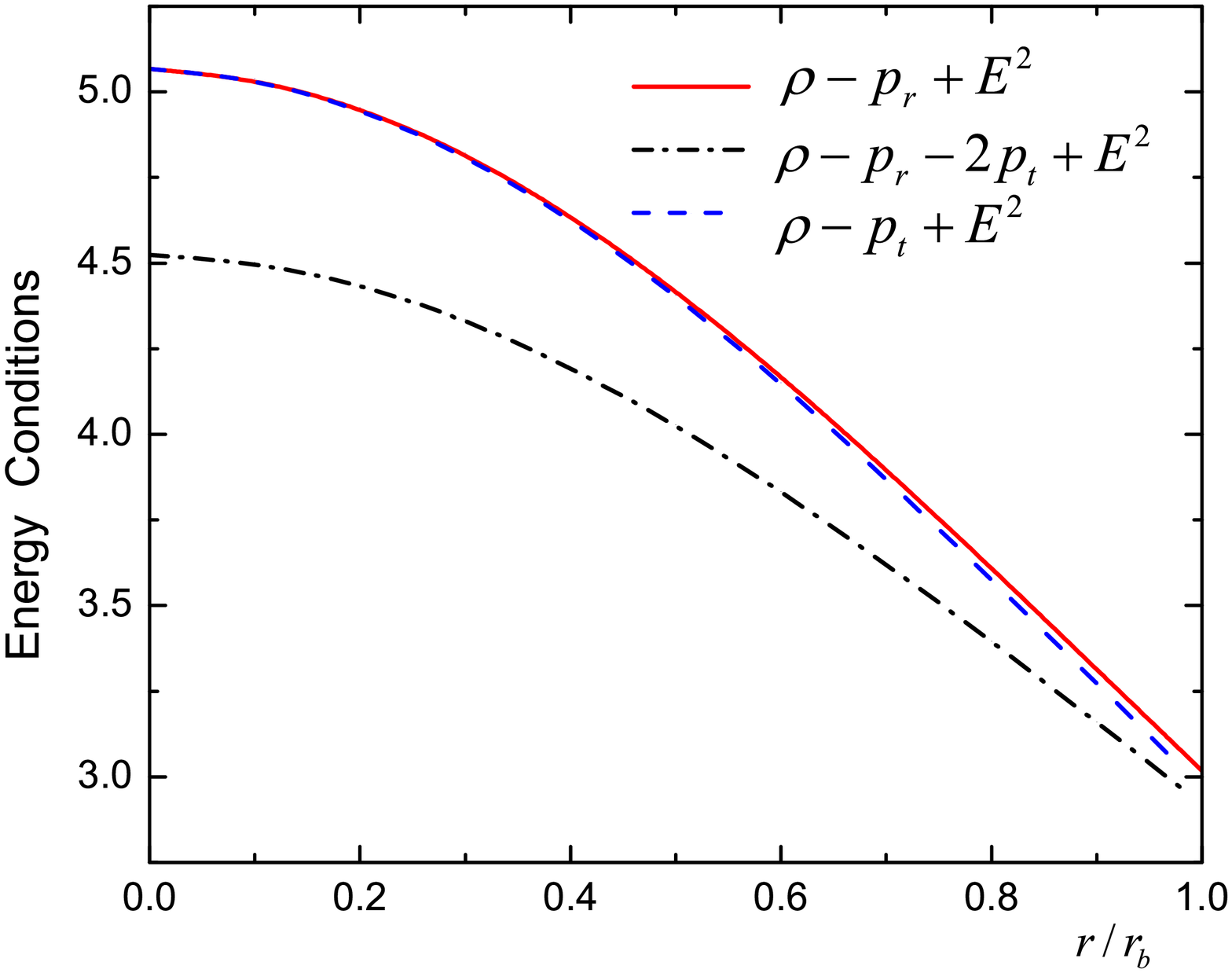}}
\caption{Energy conditions are plotted against $r$.}\label{fig13}
\end{figure}

\begin{figure}[t]
\centering
\resizebox{0.6\hsize}{!}{\includegraphics[width=4cm,height=3cm]{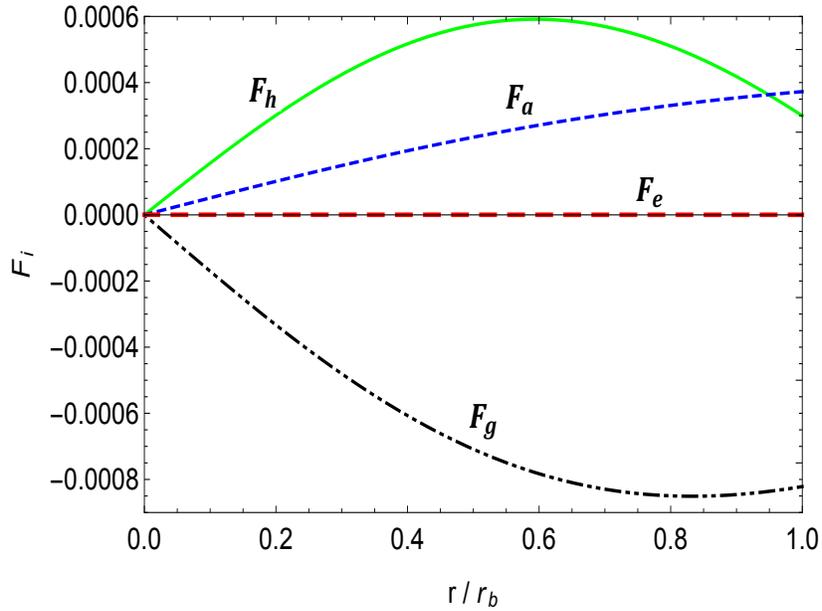}}
\caption{Forces acting on TOV-equation are plotted against $r$.}\label{fig11}
\end{figure}

On using the boundary conditions (\ref{b1})-(\ref{b3}) we get
\begin{eqnarray}
A &=& {e^X \over 1+2X} \bigg[{\alpha(2+4X)+K(2+3X)-4X-4 \over 2(1+X)} - {(1+2X)(K+\alpha-1) ~\text{Ei}(1+X) \over e^{1+X}} \bigg]\\
B &=& e^{-X} \bigg[1+AXe^{-X}-{(\alpha+K)X \over 1+X}+ {(\alpha+K-1)~\text{Ei}(1+X) \over e^{1+X}} \bigg]\\
M &=& { r_b \over 2} \bigg[1+{KX^2 \over 2(1+X)}-Be^X\bigg] \label{mb}
\end{eqnarray}
where $X=Cr_b^2$ with $r_b$ determined from surface density using (\ref{r1}). We have chosen the $K,~ \alpha$ and $X$ as free parameters.

\begin{figure}
\centering
\resizebox{0.7\hsize}{!}{\includegraphics*{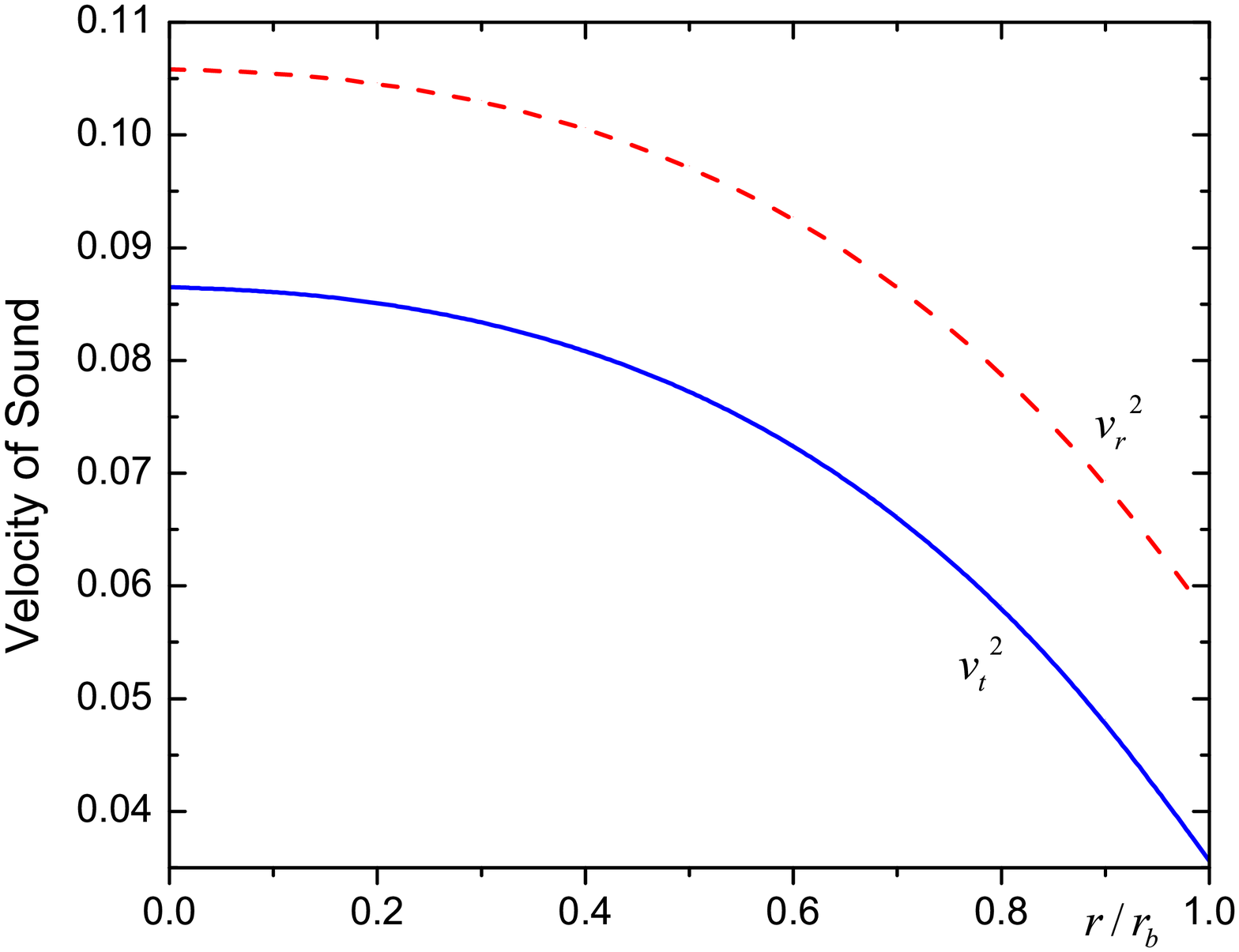}}
\caption{Square of sound speeds are plotted against $r$.}\label{fig8}
\end{figure}

\begin{figure}[t]
\centering
\resizebox{0.7\hsize}{!}{\includegraphics*{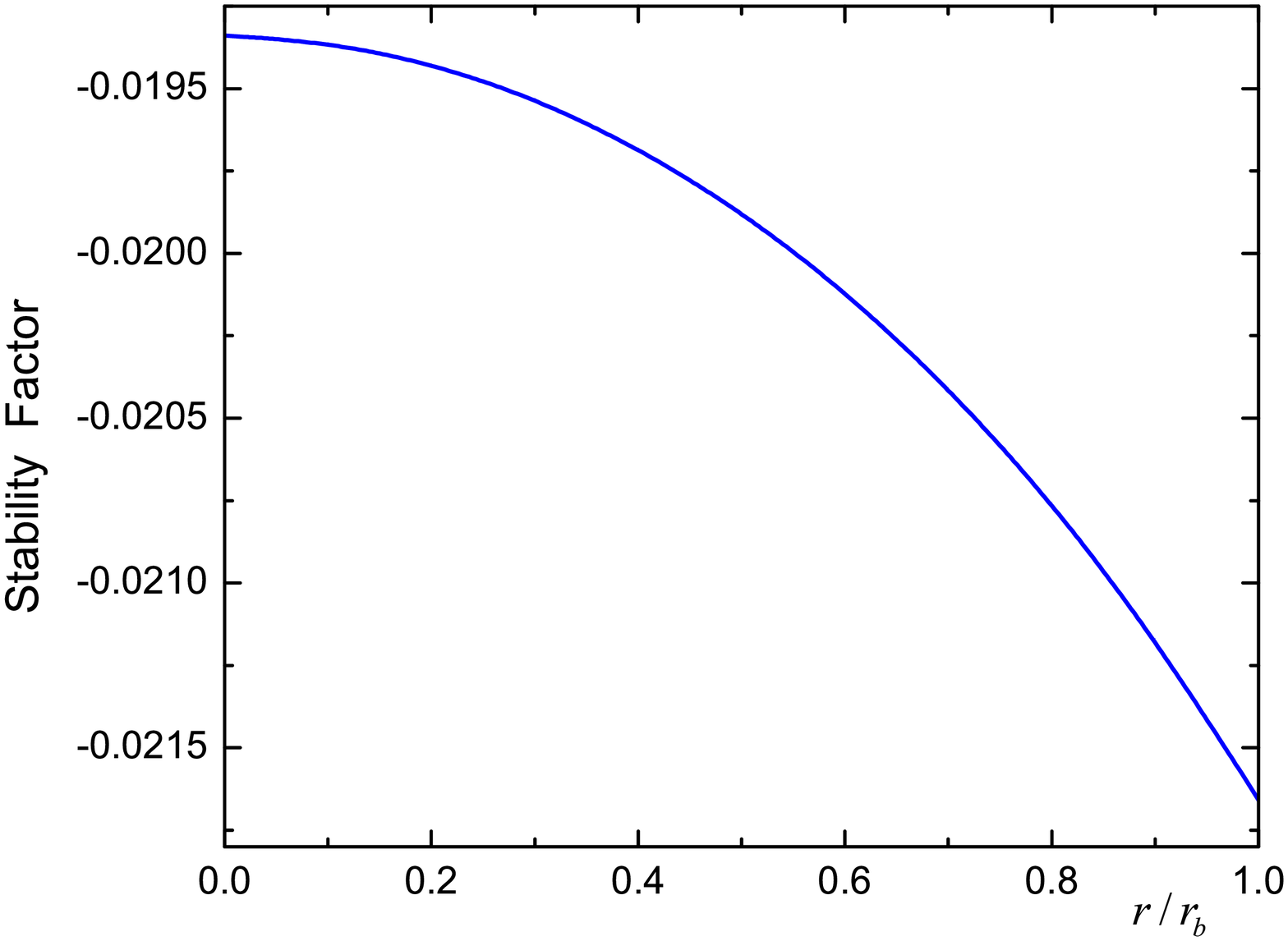}}
\caption{Stability factor is plotted against $r$.}\label{fig9}
\end{figure}

\section*{6. Mass-radius relation and compactness parameter}

The mass and compactness parameter of the compact star are obtained as,
\begin{eqnarray}
m(x) &=& \frac{2\pi}{C^{3/2}} \int \rho(x)~ x^{1/2} ~dx = \frac{1}{4 \sqrt{C}} \bigg[\sqrt{x} (2 \alpha +3 K) -2 A e^{-x} x^{3/2}-2 e^{-x-1} x^{3/2} \text{Ei}(x+1) (\alpha +K-1) \nonumber \\
&& -\frac{K}{3} x^{3/2}-\frac{2 \sqrt{x} (\alpha +K)}{x+1}-K \tan ^{-1}\left(\sqrt{x}\right) \bigg]\\
u(x) &=& {2m(x) \over \sqrt{x/C}}.
\end{eqnarray}

The profile of the mass function and compactness parameter are plotted against $r$ in fig \ref{fig12}. The profile shows that mass and compact parameter are increasing function of $r$ and they are regular everywhere inside the stellar interior.

\section*{7. Energy Conditions}

In this section we are going to verify the energy conditions namely null energy condition (NEC), dominant energy condition (DEC) and weak energy condition(WEC) at all points in the interior of a star which will be satisfied if the following inequalities hold simultaneously:
\begin{eqnarray}
\text{WEC} &:& T_{\mu \nu}t^\mu t^\nu \ge 0~~\mbox{or}~~ \rho+E^2 \geq  0,~\rho-p_i+E^2 \ge 0 \\
\text{NEC} &:& T_{\mu \nu}l^\mu l^\nu \ge 0~~\mbox{or}~~ \rho-p_i+E^2 \geq  0\\
\text{DEC} &:& T_{\mu \nu}t^\mu t^\nu \ge 0 ~~\mbox{or}~~ \rho \ge |p_i|~~ \mbox{where}~~T^{\mu \nu}t_\mu \in \mbox{nonspace-like vector} \nonumber \\
\text{SEC} &:& T_{\mu \nu}t^\mu t^\nu - {1 \over 2} T^\lambda_\lambda t^\sigma t_\sigma \ge 0 ~~\mbox{or}~~ \rho-\sum_i p_i+E^2 \ge 0. \nonumber \\
\end{eqnarray}
where $t^\mu$ and $l^\mu$ are time-like vector and null vector respectively.

We will check the energy conditions with the help of graphical representation. In Fig. \ref{fig13}, we have plotted the L.H.S of the above inequalities which verify that  all the energy conditions are satisfied at the stellar interior.

\section*{8. Stability of the model and equilibrium}

\subsection*{8.1. \it Equilibrium under various forces}
{\bf An } equilibrium state under four forces $viz$ gravitational, hydrostatics, anisotropic and electric forces can be analyzed whether they satisfy the generalized Tolman-Oppenheimer-Volkoff (TOV) equation or not and it is given by
\begin{equation}
-\frac{M_g(r)(\rho+p_r)}{r}e^{\frac{\nu-\lambda}{2}}-\frac{dp_r}{dr}+\frac{2}{r}(p_t-p_r)+\sigma E e^{\lambda/2}=0, \label{to1}
\end{equation}
where $M_g(r) $ represents the gravitational mass within the radius $r$, which can derived from the Tolman-Whittaker formula and the Einstein's field equations and is defined by

\begin{figure}[t]
\centering
\resizebox{0.7\hsize}{!}{\includegraphics*{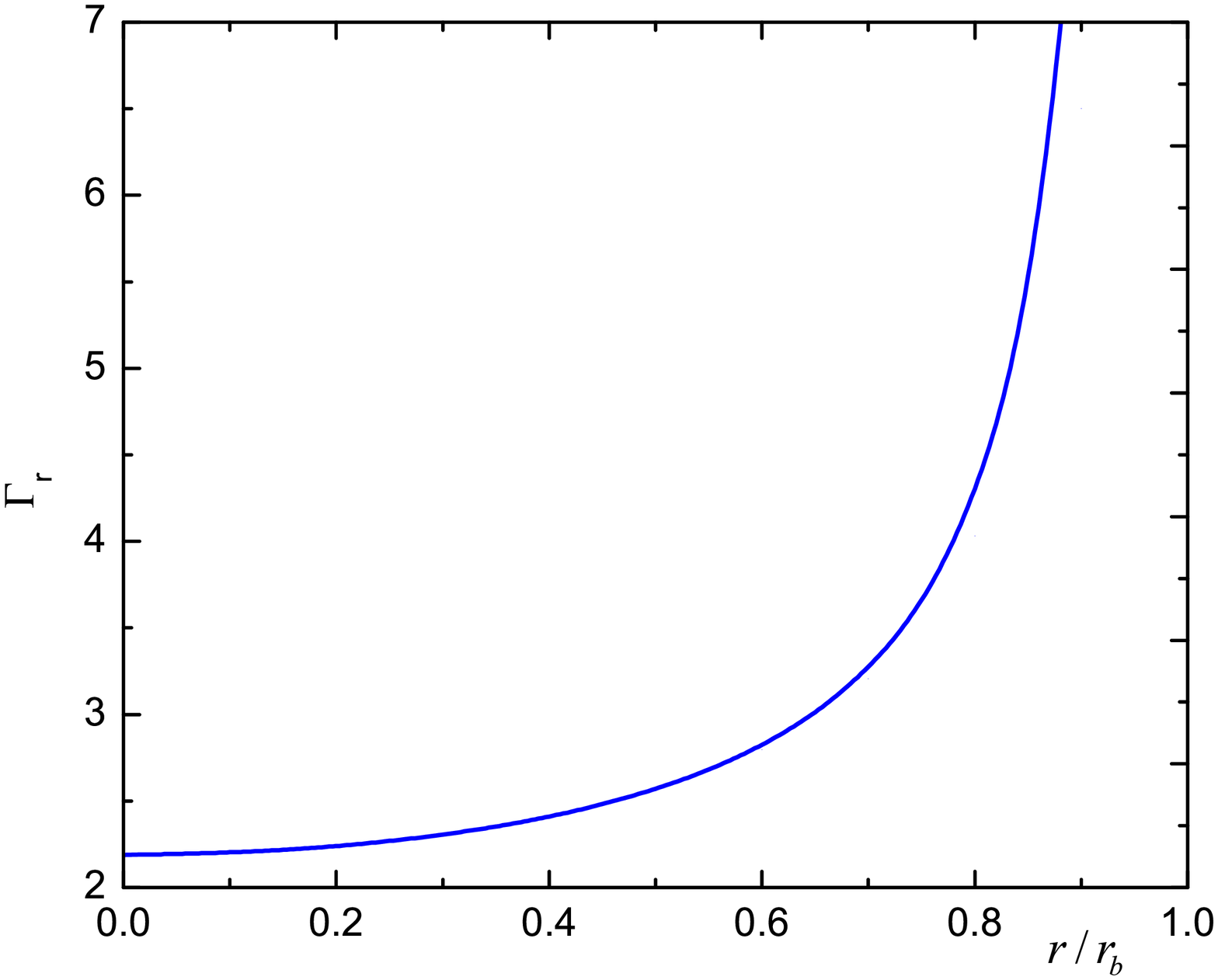}}
\caption{Adiabatic index is plotted against $r$.}\label{fig10}
\end{figure}

\begin{eqnarray}
M_g(r) &=& 4 \pi \int_0^r \big(T^t_t-T^r_r-T^\theta_\theta-T^\phi_\phi \big) r^2 e^{(\nu+\lambda)/2}dr . \nonumber \\ \label{mg}
\end{eqnarray}

For the Eqs. (\ref{dens})-(\ref{prt}), the above Eq. (\ref{mg}) {\bf reduces} to
\begin{equation}
M_g(r)=\frac{1}{2}re^{(\lambda-\nu)/2}~\nu'.
\end{equation}

\begin{figure}[t]
\centering
\resizebox{0.5\hsize}{!}{\includegraphics[width=4cm,height=3cm]{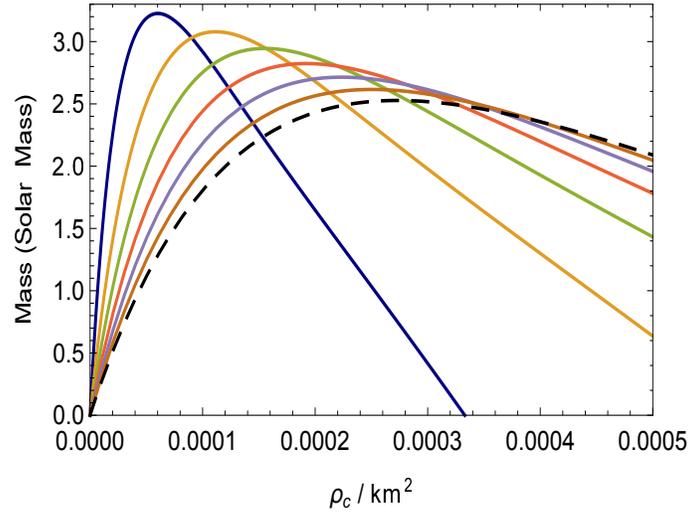}}
\caption{Variation of mass with central density for $k = 0.9 (Blue)-1.5 (Dashed),~\alpha = 0.2$ and $R=9km$.}\label{fig14}
\end{figure}

\begin{figure}[t]
\centering
\resizebox{0.5\hsize}{!}{\includegraphics[width=4cm,height=3cm]{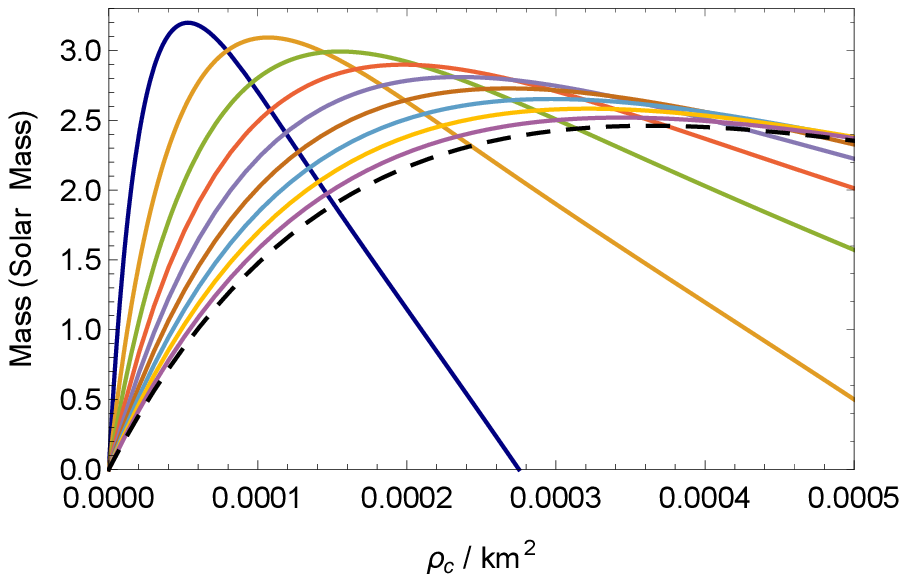}}
\caption{Variation of mass with central density is plotted for $k = 0.99,~\alpha = 0.1 (Blue)-1 (Dashed)$ and $R=9km$.}\label{fig16}
\end{figure}

Plugging the value of $M_g(r)$ in equation (\ref{to1}), we get
\begin{equation}
-\frac{\nu'}{2}(\rho+p_r)-\frac{dp_r}{dr}+\frac{2\Delta}{r}+\sigma E e^{\lambda/2}=0.
\end{equation}

The above expression may also be written as
\begin{equation}
F_g+F_h+F_a+F_e=0,
\end{equation}
where $F_g, F_h$, $F_a$ and $F_e$ represent the gravitational, hydrostatics and anisotropic and electric forces respectively.

The expressions for $F_g,~F_h$, $F_a$  and $F_e$ can be written as,
\begin{eqnarray}
F_g &=& -\frac{\nu'}{2}(\rho+p_r) = \sqrt{Cx}~\dot{\nu} (\rho+p_r),\\
F_h &=& -\frac{dp_r}{dr} = -2\sqrt{Cx} ~\dot{p}_r,\\
F_a &=& {2\Delta \over r} = {2\sqrt{C}~\Delta \over \sqrt{x}},\\
F_e &=& \sigma Ee^{\lambda/2}.
\end{eqnarray}

The profiles of three different forces are plotted in fig. \ref{fig11}. The figure shows that gravitational force is dominating is nature and is counterbalanced by the combined effect of hydrostatics, {\bf electrostatic} and anisotropic forces.

\begin{figure}[t]
\centering
\resizebox{0.7\hsize}{!}{\includegraphics*{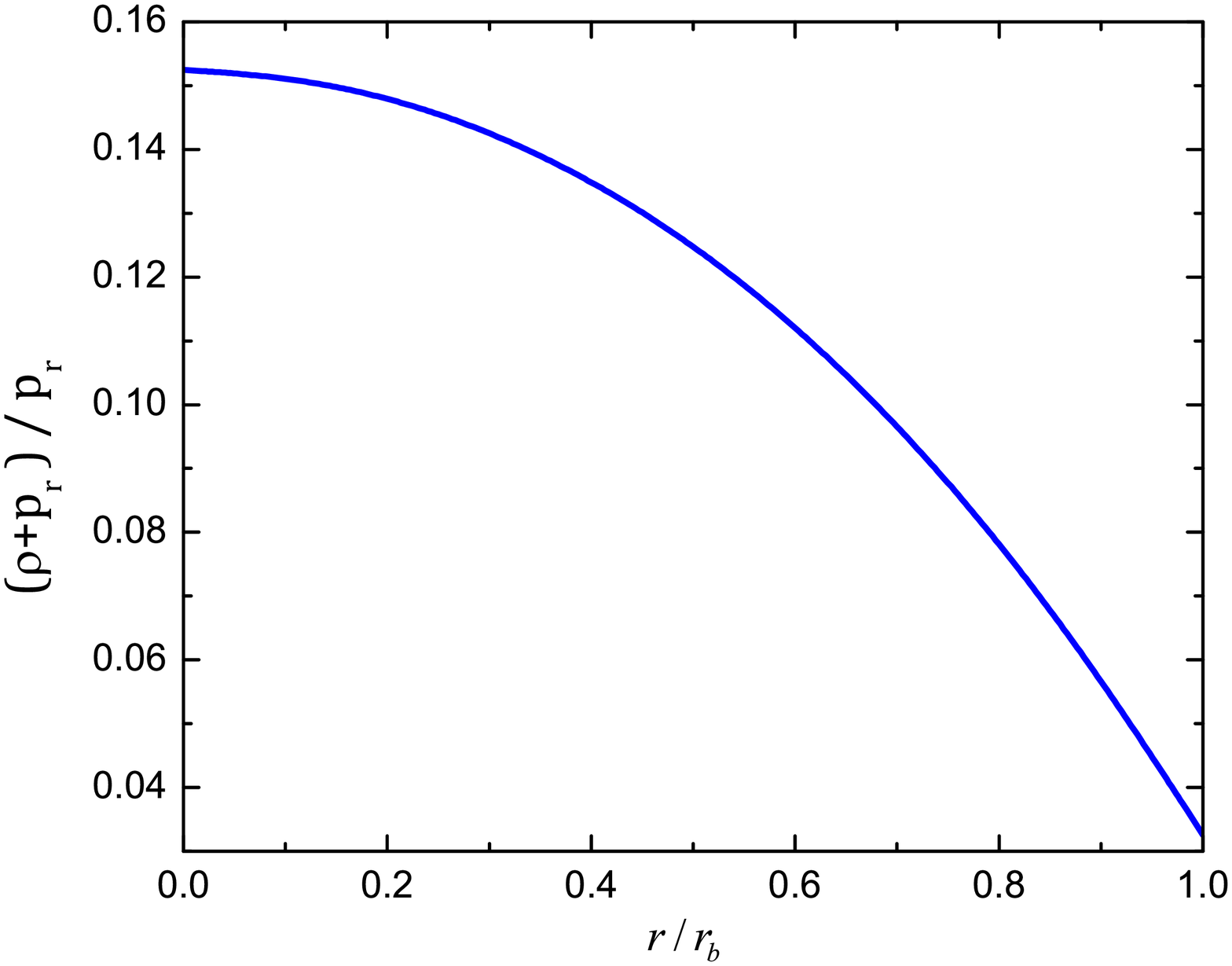}}
\caption{Ratio of trace of energy stress tensor to energy density is plotted against $r$.}\label{fig15}
\end{figure}

\begin{figure}[t]
\centering
\resizebox{0.5\hsize}{!}{\includegraphics[width=4cm,height=3cm]{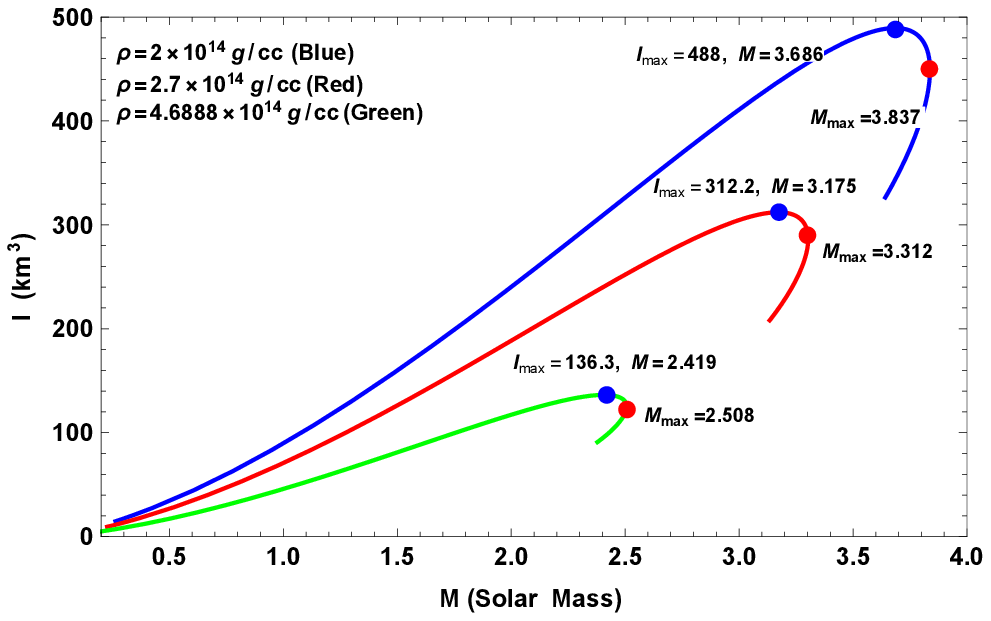}}
\caption{Variation of moment of inertia and mass for $k=0.99,~A = -0.7219$ and $\alpha = 0.2$.}\label{im}
\end{figure}

\subsection*{8.2. \it Causality and stability condition}
In this section we are going to verify the subliminal velocity of sound and stability condition. For a physically acceptable model of anisotropic fluid sphere the radial and transverse velocity of sound should be less than 1 which is known as causality condition. The radial velocity $(v_{sr}^{2})$ and transverse velocity $(v_{st}^{2})$ of sound can be obtained as

\begin{eqnarray}
v_{r}^{2} = {dp_r \over d\rho}~~,~~v_{t}^{2} = {dp_t \over d\rho}.
\end{eqnarray}

The profile of radial and transverse velocity of sound have been plotted in fig. \ref{fig8} (left), the figure indicates that our model satisfies the causality condition. Now the stability condition proposed by Abreu \cite{abr07} i.e. $0 \ge v_t^2-v_r^2 \ge -1$ is satisfied by our model (Fig. \ref{fig9} (right)).

\subsection*{8.3. \it Adiabatic index and stability condition}

For a relativistic anisotropic sphere the stability is related to the adiabatic index $\Gamma$, the ratio of two specific heats, defined by \cite{cha93},
\begin{equation}
\Gamma=\frac{\rho+p_r}{p_r}\frac{dp_r}{d\rho}.
\end{equation}

Now $\Gamma>4/3$ gives the condition for the stability of a Newtonian sphere and $\Gamma =4/3$ being the condition for a neutral equilibrium proposed by \cite{bon64}. This condition changes for a relativistic isotropic sphere due to the regenerative effect of pressure, which renders the sphere more unstable. For an anisotropic general relativistic sphere the situation becomes more complicated, because the stability will depend on the type of anisotropy. For an anisotropic relativistic sphere the stability condition is given by \cite{cha93},

\begin{equation}
\Gamma>\frac{4}{3}+\left[\frac{4}{3}\frac{(p_{ti}-p_{ri})}{|p_{ri}^\prime|r}+\frac{8\pi}{3}\frac{\rho_ip_{ri}}{|p_{ri}^\prime|}r\right]_{max},
\end{equation}
where, $p_{ri}$, $p_{ti}$, and $\rho_i$ are the initial radial, tangential, and energy density in static equilibrium satisfying (\ref{to1}). The first and last terms inside the square brackets represent the anisotropic and relativistic corrections respectively and both the quantities are positive that increase the unstable range of $\Gamma$ \cite{her92,cha93}. For this solution the adiabatic index is more than 4/3 and hence stable, Fig. \ref{fig10}.

\begin{table*}
\centering
\caption{Some well-behaved values of the parameters}
\label{tab1}       
\begin{tabular}{llllllllll}
\hline\noalign{\smallskip}
X  & $K$  & $\alpha$ & $r_b(km)$ \hspace{0.9 cm} $M / M_\odot$ & $r_b(km)$ \hspace{0.9 cm} $M / M_\odot$ &  $r_b(km)$ \hspace{1.2 cm} $M / M_\odot$ \\
&&& ($\rho_b=2\times 10^{14}g/cc$) & ($\rho_b=2.7 \times 10^{14}g/cc$) & ($\rho_b=4.6888\times 10^{14}g/cc$) \\
\noalign{\smallskip}\hline\noalign{\smallskip}
& 0 & Max=1.3 & 11.60 \hspace{1.3 cm} 0.63 & 9.60 \hspace{1.7 cm} 0.54 & 7.29 \hspace{1.7 cm} 0.41\\
0.1 & 1 & 0.57 & 11.17 \hspace{1.3 cm} 0.66 & 9.62 \hspace{1.7 cm} 0.57 & 7.30 \hspace{1.7 cm} 0.43\\
 & Max=1.75 & 0 & 11.19 \hspace{1.3 cm} 0.68 & 9.63 \hspace{1.7 cm} 0.59 & 7.31 \hspace{1.7 cm} 0.45\\
\noalign{\smallskip}\hline\noalign{\smallskip}
& 0 & Max=1.1 & 14.00 \hspace{1.3 cm} 1.37 & 12.20 \hspace{1.5 cm} 1.18 & 9.26 \hspace{1.7 cm} 0.89\\
0.2 & 1 & 0.44 & 14.13 \hspace{1.3 cm} 1.50 & 12.16 \hspace{1.5 cm} 1.29 & 9.23 \hspace{1.7 cm} 0.98\\
 & Max=1.56 & 0 & 14.15 \hspace{1.3 cm} 1.58 & 12.18 \hspace{1.5 cm} 1.36 & 9.24 \hspace{1.7 cm} 1.03\\
\noalign{\smallskip}\hline\noalign{\smallskip}
& 0 & Max=0.84 & 15.96 \hspace{1.3 cm} 2.02 & 13.74 \hspace{1.5 cm} 1.74 & 10.43 \hspace{1.5 cm} 1.32\\
0.3 & 1 & 0.33 & 15.67 \hspace{1.3 cm} 2.28 & 13.49 \hspace{1.5 cm} 1.96 & 10.23 \hspace{1.5 cm} 1.49\\
 & Max=1.42 & 0 & 15.65 \hspace{1.3 cm} 2.40 & 13.47 \hspace{1.5 cm} 2.07 & 10.22 \hspace{1.5 cm} 1.57\\
\noalign{\smallskip}\hline\noalign{\smallskip}
0.4 & 0 & Max=0.41 & 17.49 \hspace{1.3 cm} 2.62 & 15.06 \hspace{1.5 cm} 2.26 & 11.43 \hspace{1.5 cm} 1.71\\
 & Max=1.3 & 0 & 16.48 \hspace{1.3 cm} 3.11 & 14.18 \hspace{1.5 cm} 2.68 & 10.76 \hspace{1.5 cm} 2.03\\
 \noalign{\smallskip}\hline
\end{tabular}
\end{table*}

\begin{table*}
\centering
\caption{Some well-behaved values of the parameters for $\alpha=0.2$ and $K=0.5$.}
\label{tab2}       
\begin{tabular}{llllllllll}
\hline\noalign{\smallskip}
$u$  & $\rho r_b^2$  & $r_b(km)$ \hspace{1.1 cm} $M / M_\odot$ & $r_b(km)$ \hspace{1.1 cm} $M / M_\odot$ &  $r_b(km)$ \hspace{1.1 cm} $M / M_\odot$ & $z_b$ & $E_b$ \\
&& ($\rho_b=2\times 10^{14}g/cc$) & ($\rho_b=2.7 \times 10^{14}g/cc$) & ($\rho_b=4.6888\times 10^{14}g/cc$) & & \\
\noalign{\smallskip}\hline\noalign{\smallskip}
0.01 & 5.854 & 3.965 \hspace{1.3 cm} 0.026 & 3.413 \hspace{1.3 cm} 0.023 & 2.590 \hspace{1.3 cm} 0.017 & 0.010 & 0.0498\\
0.03 & 5.552 & 6.908 \hspace{1.3 cm} 0.141 & 5.945 \hspace{1.3 cm} 0.122 & 4.512 \hspace{1.3 cm} 0.092 & 0.032 & 0.0880\\
0.05 & 5.263 & 8.817 \hspace{1.3 cm} 0.299 & 7.589 \hspace{1.3 cm} 0.257 & 5.759 \hspace{1.3 cm} 0.195 & 0.054 & 0.1142\\
0.07 & 4.975 & 10.34 \hspace{1.3 cm} 0.491 & 8.899 \hspace{1.3 cm} 0.422 & 6.753 \hspace{1.3 cm} 0.320 & 0.078 & 0.1361\\
0.12 & 4.282 & 13.18 \hspace{1.3 cm} 1.070 & 11.34 \hspace{1.3 cm} 0.921 & 8.607 \hspace{1.3 cm} 0.699 & 0.144 & 0.1811\\
0.15 & 3.889 & 14.45 \hspace{1.3 cm} 1.463 & 12.44 \hspace{1.3 cm} 1.259 & 9.439 \hspace{1.3 cm} 0.955 & 0.188 & 0.2041\\
\noalign{\smallskip}\hline
\end{tabular}
\end{table*}

\subsection*{8.4. \it Harrison-Zeldovich-Novikov static stability criterion}

The stability analysis of Harrison et al. \cite{har65} and \cite{zel71} have shown that the adiabatic index of a pulsating star is same as in a slowly deformed matter. This leads to a stable configuration only if the mass of the star is increasing with central density i.e. $dM/d\rho_c > 0$ and unstable if $dM/d\rho_c < 0$.

In our solution, the mass as a function of central density can be written as
\begin{eqnarray}
M (\rho_c) &=& \frac{R}{72 \sqrt{6 \zeta  \eta }} \Bigg[ 18 \sqrt{6 \zeta  \eta }~ (2 \alpha +3 K)-16 \sqrt{6} K (\zeta  \eta )^{3/2} -96 \sqrt{6} A~ e^{\frac{3}{2}-\frac{8 \zeta  \eta }{3}} (\zeta  \eta )^{3/2}-27 K \tan ^{-1}\left(2 \sqrt{\frac{2 \zeta  \eta }{3}}\right) \nonumber \\
&& \hspace{- 0.5 cm}- \frac{108 \sqrt{6 \pi \zeta  e [e-\text{Ei}(1)]}  ~(\alpha +K-1) (\alpha +K)}{e \left(3 \alpha +3 K+8 \pi  \rho _c R^2-3\right)-3 \text{Ei}(1) (\alpha +K-1)}  -\frac{96 \pi ^{3/2} \sqrt{6 \zeta }~ \rho _c R^2 e^{\frac{1}{2}-\frac{8 \zeta  \eta }{3}} \text{Ei}\left(\frac{8 \zeta  \eta }{3}+1\right)}{[e-\text{Ei}(1)]^{3/2}} \Bigg]. \label{mrhc}
\end{eqnarray}
where
\begin{eqnarray}
\zeta = \frac{\rho _c R^2}{\alpha +K-1}~~,~~\eta = \frac{e \pi }{e-\text{Ei}(1)}. \nonumber
\end{eqnarray}
This condition can be further confirmed by Figs. \ref{fig14} and \ref{fig16}. It can also be confirmed that when the charge and anisotropy are small the range of $\rho_c$ for $dM/d\rho_c>0$ is very narrow and for larger values of charge and anisotropy we get wider range of $\rho_c$ for $dM/d\rho_c>0$. This implies that the static stable configuration of a compact star with very less electric charge and anisotropy may be altered by small radial perturbations/oscillations. However,  inclusion of more electric charge and anisotropy can enhance the stability of compact stars under small perturbations.

\section*{9. Moment of inertia and EoS}

For a uniformly rotating star with angular velocity $\Omega$, the moment of inertia is given by \cite{latt}
\begin{eqnarray}
I = {8\pi \over 3} \int_0^R r^4 (\rho+p_r) e^{(\lambda-\nu)/2} ~{\omega \over \Omega}~dr
\end{eqnarray}

where, the rotational drag $\omega$ {\bf satisfies} the Hartle's equation \cite{hart}
\begin{eqnarray}
{d \over dr} \left(r^4 j ~{d\omega \over dr} \right) =-4r^3\omega~ {dj \over dr} .
\end{eqnarray}
with $j=e^{-(\lambda+\nu)/2}$ which has boundary value $j(R)=1$. The approximate solution of moment of inertia $I$ up to the maximum mass $M_{max}$ was given by Bejger and Haensel \cite{bejg} as
\begin{equation}
I = {2 \over 5} \Big(1+x\Big) {MR^2},
\end{equation}
where parameter $x = (M/R)\cdot km/M_\odot$. For {\bf the explored solution,} we have plotted mass vs $I$ in Fig. \ref{im} {\bf which} shows that if $n$ increases, the mass increases and the moment of inertia increases till up to certain value of mass and then decreases. Comparing Figs. \ref{mrn} and \ref{im} we can see that the mass corresponding to $I_{max}$ is not equal to $M_{max}$ from $M-R$ diagram. In fact the mass corresponding to $I_{max}$ is lower by  $\sim 3.55$\% from the $M_{max}$. This happens to the EoSs without any strong high-density softening due to hyperonization or phase transition to an exotic state \cite{bej}. Using this graph we can estimate the maximum moment of inertia for a particular compact star or by matching the observed $I$ with the $I_{\max}$ we can determine the validity of a model.

\section*{10. Results and Discussions}

We have explored a well-behaved charge analogue of anisotropic Kuchowicz solution. The solution is well behaved for $0< X \le 0.6$, $0 \le \alpha \le 1.3$, $0< K \le 1.75$ and Schwarzschild compactness parameter $0< u \le 0.338$. Since our solution is well behaved for a wide ranges of different parameters, one can use to model many different types of ultra-cold compact stars i.e. quark stars and neutron stars.

From table \ref{tab1}, it is observed that the increase in charge parameter results to increase in maximum mass, however, increase in anisotropy results in decrease in maximum mass. {\bf With the} increase in charge the Columbic force enhances the outward pressure that can support more mass while increase in anisotropy diverts more pressure away from radial direction thereby decreasing the mass.

In table \ref{tab2}, we present some models of super dense quark stars and neutron stars corresponding to $X=0.2,$ $\alpha =0.2$ and $K=0.5$ for which $0.01 \le u \le 0.15$. For surface density $\rho_b=4.6888 \times10^{14} g/cc$ the range of mass is $0.017 \le M_{max} \le 0.955$ and radius $2.59 km \le r_{b} \le 9.439 km$ . For $\rho_b=2.4 \times10^{14} g/cc$ the range of mass and radius are  $0.023 \le M_{max} \le 1.259$ and $3.493 km \le r_{b} \le 12.44 km$ respectively. For $\rho_b=2 \times10^{14} g/cc$ the range of mass and radius are  $0.026 \le M_{max} \le 1.463$ and $3.965 km \le r_{b} \le 14.45 km$  respectively, {\bf which shows the robustness of our present work in astrophysical scenario.}

Furthermore, our presented solution satisfies Weak Energy Condition (WEC), Null Energy Condition (NEC) and Dominant Energy Condition (DEC). The static stability criterion i.e. $dM/d\rho_c > 0$, $\Gamma > 4/3$ and $-1 \le v-t^2-v_r^2 \le 0$ guaranteed that the presented solution is static and stable. Also, the satisfaction of TOV-equation implies that the solution can represent configurations at equilibrium. Furthermore, the ratio of trace of energy-momentum tensor is less than unity i.e. $(\rho+p_r)/p_r \le 1$ (see Fig. \ref{fig15}) signifies that the solution can describe physical matters.

Another important conclusion is that for larger the values of anisotropy and electric charge the stability of compact stars is enhanced and therefore remains static stable under radial oscillations. Unlike the other calculations, here we have assumed the surface density of the compact stars and the radius is determined from it, then by using the radius we determine the maximum mass of the configuration applying boundary condition  (\ref{mb}). Hence, the radius and mass of the configuration presented in the model are surface density dependent. The M-R diagram in Fig.(\ref{mrn}) clearly suggests that the maximum mass of the stellar configurations strongly depend on the surface density. Here we have plotted the graphs for neutron star with surface densities $2\times $10$^{14}~g/cc$ and $2.5 \times 10^{14}~g/cc$ and for quark star with surface density $4.6888 \times $10$^{14}~g/cc$. These suggest that more the surface density of the compact star lesser is the maximum mass it can hold. This may be because of the fact that at lower densities the equation of state (EoS) is stiffer when the dominating particles are mainly neutrons, however, as density increases many other particles such as hyperons \cite{cam,amb,pan}, pion (condensate) \cite{web}, quarks etc. are generate and consequently soften the EoS thereby decreasing the maximum mass.

Herrera et al. \cite{her08} have shown that all the spherically symmetric solutions of field equations can always be generated by two generators $\zeta(r)$ and $\Pi(r)$ defined as
\begin{eqnarray}
e^{\nu(r)} &=& \exp \left[\int \left\{2\zeta(r)-{2 \over r} \right\}~dr\right] \\
\Pi(r) & = &8\pi (p_r-p_t).
\end{eqnarray}

and these generators are
\begin{eqnarray}
\zeta(r) &=& B C r ~e^{C r^2}+\frac{1}{r}\\
\Pi(r) &=& -\Delta(r).
\end{eqnarray}

It is interesting to note that among two generating functions, one is positive  and other is negative and  both are increasing in nature. That means theoretically we can predict physically viable stars   by choosing generating functions.  So the   investigation of  the nature of the generating functions will be an active area of research in near future. 

\section*{11. Conclusion}
All the above rigorous analysis implies that the solution is suitable to represent compact star solution which might be useful in future.

\section*{Acknowledgement}
FR would like to thank the authorities of the Inter-University Centre for Astronomy and Astrophysics, Pune, India for providing research facilities. FR is  also grateful to
DST-SERB (Grant No.: EMR/2016/000193) , Govt. of India, for financial support.


\end{document}